\documentclass[12pt]{elsarticle}
\usepackage[margin=1in]{geometry}
\usepackage{braket}
\usepackage{subcaption}
\usepackage{upgreek}
\usepackage{physics}
\usepackage{amssymb}
\usepackage{array}

\begin{document}

\title{Collective motion from quantum-inspired dynamics in visual perception}

\author{
Jyotiranjan Beuria$^{1,3}$, Mayank Chaurasiya$^{1}$ and Laxmidhar Behera$^{1,2}$}

\address{$^{1}$Indian Institute of Technology Mandi, Himachal Pradesh, India\\
$^{2}$Indian Institute of Technology Kanpur, Uttar Pradesh, India\\
$^{3}$IKS Research Centre, ISS Delhi, Delhi, India
}

\begin{abstract}
We propose a model of collective behavior in self-propelled active agents that incorporates a perceptual decision-making process. In this framework, the decision-making dynamics is modeled using quantum formalism. The perceptual decision state of each agent is an entangled or superposed state of the decision states for the neighboring agents within the vision cone. We suggest that in this framework, the force driving the movement of active agents is governed by the quantum average of its perception operator, providing a bridge between perceptual decision-making processes and classical dynamics. Additionally, we introduce two perceptual measures of cohesion in the flock, namely, perception strength and perceptual energy, to characterize collective behavior in terms of decision-making dynamics. Our model demonstrates that, with an appropriate choice of perceptual decision state, the well-known Vicsek model of flocking behavior can be derived as a specific case of this quantum-inspired approach. This approach provides fresh insights into collective behavior and multi-agent coordination, revealing how classical patterns of collective behavior emerge naturally from perception. 
\end{abstract}




\maketitle

\section{Introduction}
The emergence of order from the collective behavior \cite{sumpter2010collective,vicsek2012collective,popkin2016physics} of self-propelled agents is ubiquitous in Nature. The universality of spectacular coordinated behavior has been observed at very different sizes and scales, such as the flocking of birds \cite{ballerini2008interaction}, schooling of fish \cite{rosenthal2015revealing}, bacterial colonies \cite{rabani2013collective}, locust swarms \cite{buhl2012using}, sheep herds \cite{gomez2022intermittent}, and even human crowding \cite{ma2021spontaneous,keta2022disordered} or robot swarming \cite{zhao2018self}. One of the primary characteristics of the models describing this collective behavior is the emergence of long-range velocity correlations. Such correlations result when attractive, aligning, and repulsive interactions are neatly balanced. 

Animals engaging in coordinated group movement, such as birds, fish, mammals, and insects, depend heavily on vision to interpret the behavior of nearby individuals and align their actions accordingly. Vision-based decision-making that leads to collective behavior is equally crucial for artificial agents, particularly in the rapidly developing domain of swarm robotics. While this vision-based interaction is fundamental to the emergence of collective behavior, the specific perceptual and cognitive mechanisms behind it remain inadequately understood. The challenge is particularly acute in visually crowded environments, such as in a flock, where distinguishing relevant cues is difficult. Moreover, animals must make these decisions efficiently, despite possessing limited neural and computational resources \cite{sumpter2006principles}. This constraint necessitates streamlined processing strategies that can rapidly extract essential motion cues like the direction, spacing, and configuration of neighbors from noisy or ambiguous sensory input.

Recent studies, such as those on desert locusts \cite{bleichman2023visual,bleichman2024visual}, suggest that animals prioritize particular features, such as the count of moving elements or the structure of their motion, rather than processing all visual data equally. This highlights an adaptive and efficient balance between available perceptual resources and behavioral effectiveness in guiding collective motion. To date, modeling approaches for flocking that incorporate both local and non-local interactions \cite{beuriaricci2024} based on visual perception have predominantly employed classical nonlinear dynamical systems driven by visual perceptual inputs. However, this work attempts to model the underlying cognitive decision-making leading to the physical motion of these self-propelled agents through the lens of quantum dynamics.

In recent years, there has been a surge of interest in exploring cognition \cite{khrennikov2015quantum,ashtiani2015survey,broekaert2017quantum,pothos2022quantum,khrennikov2023open}, particularly decision-making \cite{asano2012quantum,asano2017quantum,meghdadi2022quantum,khrennikov2022quantum}, through the mathematical framework of quantum mechanics. These approaches are referred to as quantum-like or quantum-inspired because they draw formal parallels with quantum systems without necessarily implying the presence of actual quantum processes in the brain \cite{hameroff2014consciousness}. Importantly, quantum-like cognitive models \cite{khrennikov2003quantum,broekaert2017quantum,asano2017quantum,li2020quantum,meghdadi2022quantum} are distinct from theories that propose a physical quantum brain. Whereas traditional quantum states are described by wavefunctions in physical space-time, these quantum states are defined in an abstract cognitive space. This serves as an operational mathematical construct to capture the dynamics of cognitive processes. In recent years, there have been attempts to ground quantum-like cognitive states with neurophysiological processes \cite{khrennikov2018quantum} and through the consideration of emergent states of $k$-regular graphs \cite{scholes2024quantum,scholes2025quantumlike} of neuronal oscillators.

There is substantial evidence indicating that human decision-making often deviates from the predictions of classical probability theory \cite{khrennikov2015quantum,meghdadi2022quantum}.
In quantum-inspired approaches to modeling cognition, the notion of superposition is especially important for decision-making, as it captures states of profound uncertainty where individuals hold competing preferences or judgments simultaneously, something classical probability cannot adequately represent \cite{khrennikov2015quantum}. Moreover, the use of non-commuting operators to model incompatible cognitive observables aligns well with the order effects \cite{trueblood2011quantum} often observed in human decisions, where the sequence of options can change outcomes. Entanglement, central to many quantum phenomena, also plays an important role in modeling decision-making by reflecting the contextual interdependence of choices, attitudes, or beliefs \cite{meghdadi2022quantum,khrennikov2023entanglement}. Rather than invoking physical non-locality, entanglement in these quantum states points to the way decisions are shaped by the shared surrounding cognitive and social context. A similar departure from classical probabilistic principles is also likely to underlie the flocking behavior of birds and animals. Given the limited neural resources available to birds, a quantum-inspired framework offers a promising and biologically plausible approximation of their underlying cognitive mechanisms, providing a more compelling alternative to classical nonlinear dynamical models. It is worth mentioning that there have been recent studies reporting the advantages of quantum-inspired algorithms over their classical counterparts \cite{mohseni2022ising,alodjants2024quantum}.

Although there have been several works in quantum-like modeling of human decision-making, its application for the collective behavior of self-propelled agents, such as the flocking of birds, has not captured much attention. That is precisely the scope of this work. As mentioned, vision is a key perceptual mechanism for birds to form cohesive patterns. Physical space momenta, and position define the physical states of these self-propelled active agents. We define the perceptual decision states of these agents, which are distinct from their physical states. By introducing a so-called perception operator, we also reproduce the popular Vicsek dynamics \cite{vicsek1995novel} in the physical space as a special case of this model. Although we develop this model keeping the flocking of birds in mind, it can be operationally extended to any multi-agent system with a generic notion of perception of neighbors.

The organization of this paper is as follows. In section \ref{sec:framework}, we introduce the quantum-inspired framework used to define the perceptual decision states of active agents. Section \ref{sec:perception-op} presents the formulation of a Hermitian perception operator and its evolution, giving rise to collective behavior in physical space. Section \ref{sec:vicsek} demonstrates how the proposed perceptual dynamics naturally recovers the key structure of the Vicsek model for flocking. We also present numerical simulations in section \ref{sec:numerical}. Finally, we conclude and discuss future directions in section \ref{sec:conclusion}.

\section{The Framework}
\label{sec:framework}

\begin{figure}[!ht]
    \centering
    \includegraphics[width=\linewidth]{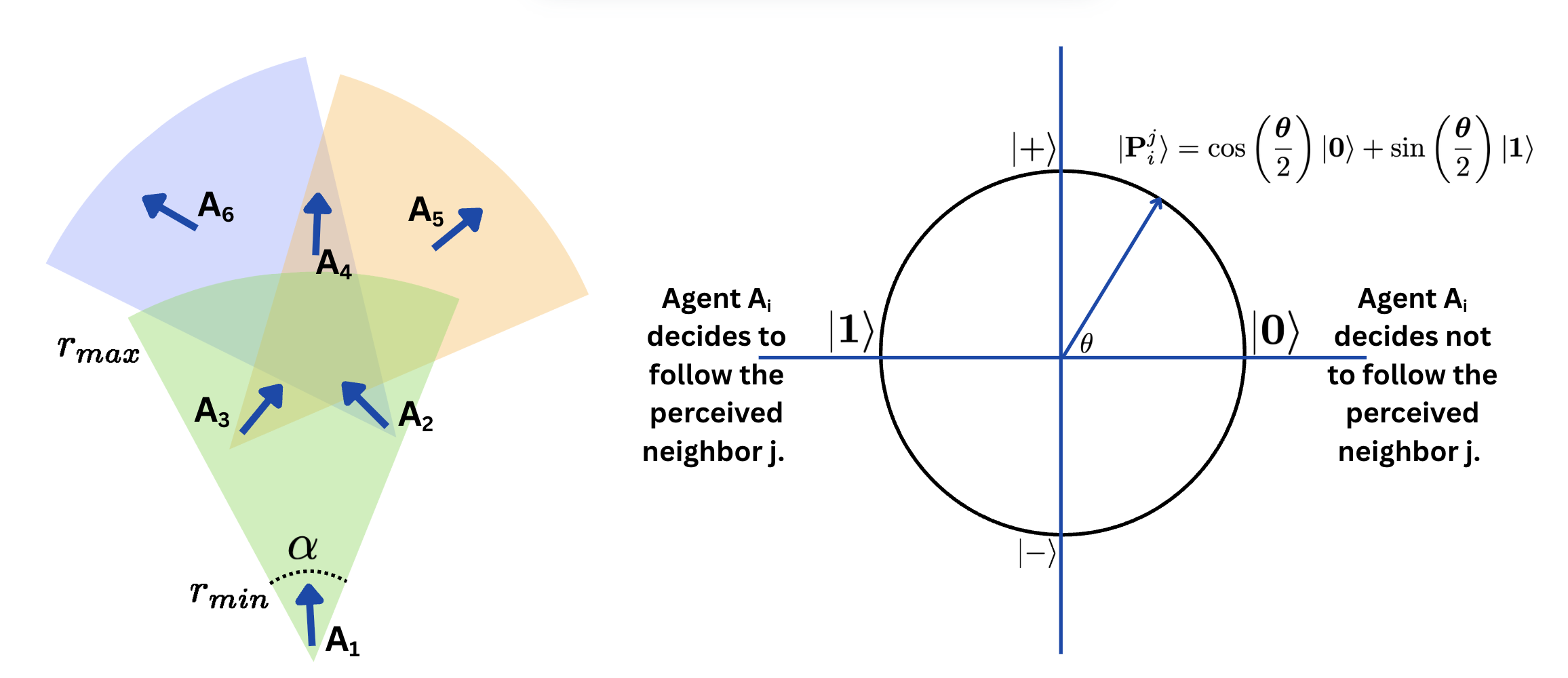}
    \caption{An agent $\text{A}_i$ perceives few neighbors $\text{A}_j$ in its vision cone defined by angular width $\alpha$ along the direction of motion and constrained to a radial range between $r_{\min}$ and $r_{\max}$. The perceptual decision state of $\text{A}_i$ in relation to neighbor $\text{A}_j$ is denoted as a superposed state $\ket{\text{P}_i^j}$.}
    \label{fig:entanglement}
\end{figure}

Consider a system of \( N \) self-propelled active agents of unit mass moving at a constant speed that reach consensus and forms a cohesive flock. The collective motion can essentially be described in terms of a decision-making process of following a neighbor or not. The agents lying within the vision cone are called neighbors. We define the vision cone by an angular width $\alpha$ centered along the direction of motion and constrained to a radial range between $r_{\min}$ and $r_{\max}$ (see Figure \ref{fig:entanglement}). We propose a quantum-inspired formalism for representing such a perceptual decision-making process. 

We consider the simplest case of an agent  $\text{A}_i$ that perceives a single neighbor $\text{A}_j$ in its vision cone. Let $\ket{\text{P}_i^j}$ represent the decision state of the agent  $\text{A}_i$  as it perceives its neighbour $\text{A}_j$. The agent $\text{A}_i$ has the orthogonal binary option of either following $\text{A}_j$ or not following $\text{A}_j$. Let this orthogonal decision basis that encodes binary decisions be $\ket{1}$: to follow  and $\ket{0}$: not follow.  $\ket{\text{P}_i^j}$ is the state of the agent $\text{A}_i$ in the decision space, which can be likened to the superposition of either to follow the neighbor or not and thus expressed as:  
\begin{equation}
\ket{\mathrm{P}_i^j} = \cos\left(\frac{{\theta}_{i}^j}{2}\right)\ket{{0}}_i^j + \sin\left(\frac{{\theta}_{i}^j}{2}\right)\ket{{1}}_i^j,
\end{equation}
where \( \theta_{i}^j \) encodes the subjective evaluation of agent \( \text{A}_i \) towards following agent \( \text{A}_j \). The proposition of a superposed state is motivated by the fact that the agent is simultaneously aware of both choices.

Let us consider a case wherein an agent within the flock perceives only two neighboring agents. 
The agent has to choose one of the following four possible decision combinations: (1) follow both neighbors, (2) follow neither, (3) follow the first but not the second, and (4) follow the second but not the first. These four elementary decision-making choices can be represented as $\ket{11}$, $\ket{00}$, $\ket{10}$, and $\ket{01}$, respectively. This forms the orthogonal basis states for the two-neighbor case. 

The agent's decision-making process is assumed to be concurrently aware of all four available options. Thus, the generic decision state can be expressed as the superposition of these four basis choices. We denote such a perceptual decision-making state of $\text{A}_i$ with $\ket{\mathrm{P}_i}$, which is in turn conceptually distinct from its classical physical state. Mathematically, we can write as follows:
\begin{equation}
    \ket{\mathrm{P}_i} = a_i\ket{00}_i + b_i\ket{01}_i + c_i\ket{10}_i  + d_i\ket{11}_i, 
    \label{eq:random_sup}
\end{equation}
where the basis states are specific to the neighborhood of $\text{A}_i$. A special case of such a superposed state is called uniform superposition, wherein all four elementary choices mentioned earlier are equally viable. Mathematically, it refers to $a_i=b_i=c_i=d_i=\frac{1}{2}$. Such a uniform superposition state is a separable state. It essentially implies that agent $\text{A}_i$ makes decisions about each of the observed neighbors independently, without considering joint or correlated interactions. It can be expressed as: 
\begin{align}
\ket{\mathrm{P}_i} & = \ket{\mathrm{P}_i^1} \bigotimes \ket{\mathrm{P}_i^2}\\
&= \frac{1}{\sqrt{2}}\left(\ket{0}_i^1+\ket{1}_i^1\right) \bigotimes \frac{1}{\sqrt{2}}\left(\ket{0}_i^2+\ket{1}_i^2\right).
\end{align}
However, this may not be the case always. In a generic scenario, when an agent's decision-making for the first neighbor is contingent upon the second one, we term such a state as entangled. In other words, $\ket{\mathrm{P}_i} \neq \ket{\mathrm{P}_i^1} \bigotimes \ket{\mathrm{P}_i^2}$. Thus, for a random choice of coefficients in equation \ref{eq:random_sup}, only those combinations satisfying $a_i d_i = b_i c_i$ \cite{jorrand2003separability} correspond to a separable (purely superposed) state. In all other cases, the resulting state is entangled, indicating that the perceptual decisions regarding the two neighbors are interdependent rather than independent. 

Let's consider some popular choices of entangled states now. In one such scenario, the agent simultaneously decides either to follow both neighbors or to follow neither. When both these choices are equally probable, we can express it as follows:
\begin{equation}
    \ket{\mathrm{P}_i} =\ket{\Phi_i^+} = \frac{1}{\sqrt{2}} (\ket{00}_i + \ket{11}_i)
    \label{eq:bell1}
\end{equation}
When the agent chooses to follow only one of the two neighbors at a time while simultaneously not following the other, the corresponding perceptual decision state can be represented as
\begin{equation}
\ket{\mathrm{P}_i} =\ket{\Psi_i^+} = \frac{1}{\sqrt{2}} \left( \ket{01}_i + \ket{10}_i \right),
\label{eq:bell2}
\end{equation}
indicating a superposition of mutually exclusive decision outcomes. Such states represent mutually interdependent decision states that cannot be factored in terms of independent individual decisions. The states in equations \ref{eq:bell1} and \ref{eq:bell2} are called the maximally entangled Bell states.

There can be a scenario when the possible elementary choices give rise to mutual perceptual conflict, anti-aligned evaluations, and destructive interference. Such perceptual decision states can be represented by some negative coefficients in both purely superposed states and entangled states.  Our numerical simulation reveals that such scenarios do not give rise to cohesion in the flock. In case of Bell states mentioned earlier, these anti-aligned perceptual decision states are represented by $\ket{\Psi_i^-}= \frac{1}{\sqrt{2}} \left( \ket{01}_i - \ket{10}_i \right)$ and $\ket{\Phi_i^-}=\frac{1}{\sqrt{2}} \left( \ket{00}_i - \ket{11}_i \right)$.

We consider the case where an agent perceives three neighbors within its vision cone. In the generic case, the state of the agent $\text{A}_i$ in its decision space is the superposition of all eight basis states:
\begin{equation}
\ket{\text{P}_i}=a_i \ket{000} + b_i \ket{001} + c_i \ket{010} + d_i \ket{011}+ e_i \ket{100} + f_i \ket{101} + g_i \ket{110} + h_i \ket{111}  
\label{eq:three_agents}
\end{equation}
In a possible case where the decision of one neighbor is contingent upon the other two, such a state is an entangled state. For example, if an agent either follows all three neighbors or none at all or follows exactly one of the three. These decision-making configurations correspond to entangled states and can be expressed using the Greenberger-Horne-Zeilinger (GHZ) state and the W state, respectively, as follows:
\begin{align}
    \ket{\text{P}_i} &= \ket{\text{GHZ}_3}= \frac{1}{\sqrt{2}} \left(\ket{000} + \ket{111}\right), \nonumber \\
    \ket{\text{P}_i} &=\ket{\text{W}_3} = \frac{1}{\sqrt{3}} \left( \ket{001} + \ket{010} + \ket{100} \right).
    \label{eq:trace_wghz}
\end{align}

Along similar lines, the state of an agent with four neighbors can be expressed as a superposition of 16 possible basis states. Mathematically:
\begin{align}
    \ket{\text{P}_i} &= \ket{\text{GHZ}_4}= \frac{1}{\sqrt{2}} \left(\ket{0000} + \ket{1111}\right), \nonumber \\
    \ket{\text{P}_i} &=\ket{\text{W}_4} = \frac{1}{\sqrt{3}} \left( \ket{0001} + \ket{0010} + \ket{0100} +\ket{1000} \right).
    \label{eq:trace_wghz_4}
\end{align}
We also present a special class of entangled states, called the cluster states, in the Appendix section \ref{sec:trace}. For simplicity, we restrict ourselves to two and three-neighbor cases for subsequent numerical simulations. 

It is worth mentioning that as the physical state of the system evolves, the distribution of neighbors within the vision cone of an agent also changes. However, the binary decision basis remains the same (to follow or not to follow). Thus, we can imagine these decision states for neighbors as some kind of template states that correspond to different physical attributes at every time step. The empirical finding on the optimal number of neighbors in the flock of starlings has been found to be seven \cite{ballerini2008interaction}. Our previous theoretical analysis \cite{beuriaricci2024} also shows that even one neighbor can ensure the stability of the flocking behavior. These findings also support our choice of a few neighbors in our subsequent numerical results.


\section{The Perception Operator}
\label{sec:perception-op}

We describe a notion of a perception operator, which is crucial for connecting the perceptual decision-making dynamics with physical space dynamics. In subsequent sections, we will also derive the Vicsek model \cite{vicsek1995novel}, commonly used to study the dynamics of flocking agents.

Let $\rho_i$ be the density matrix at time $t$ representing an agent in its so-called perceptual Hilbert space $\mathcal{H}_i$ formed by the $n$ randomly selected neighbors through the scheme described in Figure \ref{fig:entanglement}. 
We now define the perception operator $O_i^k$ acting on $\mathcal{H}_i$ for the $k$-th spatial direction in a $d$-dimensional periodic box. We will show later that the off-diagonal elements of the perception operator contribute to the alignment, and the diagonal elements contribute to the noise or uncertainty in perceptual dynamics. For $n$ neighbors, we introduce a hermitian operator for $O_i^k$ of dimension $\mathcal{N}=2^n$ as follows.
\begin{align}
    O_i^k &= \frac{1}{n}\begin{pmatrix}
    n\eta_i^k & u_{12} e^{i \phi_{12}} & \cdots & u_{1\mathcal{N}} e^{i \phi_{1\mathcal{N}}} \\
    u_{21} e^{i \phi_{21}} & n\eta_i^k  & \cdots & u_{2\mathcal{N}} e^{i \phi_{2\mathcal{N}}} \\
    \vdots & \vdots & \ddots & \vdots \\
    u_{\mathcal{N} 1} e^{i \phi_{\mathcal{N} 1}} & u_{\mathcal{N} 2} e^{i \phi_{\mathcal{N} 2}} & \cdots & n\eta_i^k
    \end{pmatrix},
    \label{eq:o_matrix}
\end{align}
where $u_{\alpha \beta}=u_{\beta \alpha}$ is a real number and $\phi_{\alpha \beta}=-\phi_{\beta \alpha}$ is the phase angle. The diagonal elements $\eta_i^k$ are constant factors contributing to the noise in the perception. Throughout the discussion, we shall keep a small nonzero constant value for $\eta_i^k=\eta$, except when we are studying the phase transition in terms of the order parameter. 

\begin{figure}[!ht]
    \centering
    \includegraphics[width=0.5\linewidth]{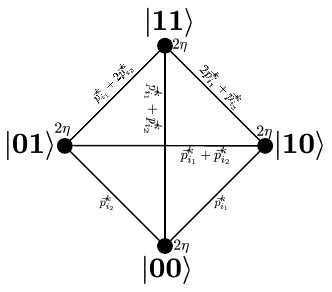}
    \caption{Representation of perception operator for the two neighbors case as a weighted graph (up to an overall factor of $\frac{1}{n}$).}
    \label{fig:graph_op}
\end{figure}

We further define
\begin{align}
    \phi_{\alpha \beta}&=0,\\
    u_{\alpha \beta} &= (\vec{L}_\alpha + \vec{L}_\beta)\cdot \vec{{\Gamma}}_i^k,
\end{align}
where $\vec{L}_\alpha$ and $\vec{L}_\beta$ are the $n$-dimensional label vectors of the matrix and $\vec{\Gamma}_i^k$ is the vector formed by $k$-th momentum component of $n$ neighbors. Physically, the off-diagonal ($\alpha$, $\beta$) component of  $O_i^k$ quantifies the transition and coherence between $\ket{\alpha}$ and $\ket{\beta}$ under its action. This perception operator resembles the structure of a weighted adjacency matrix. Thus, it can be represented as a fully connected graph. For $n=2$, the perception operator (up to $\frac{1}{2}$ factor) becomes as follows.

\begin{equation}
\begin{array}{c|*{4}{c}}
     & \ket{00} & \ket{01} & \ket{10} & \ket{11} \\
\hline
\ket{00} & 2\eta & \vec{p}_{i_2}^k & \vec{p}_{i_1}^k & \vec{p}_{i_1}^k + \vec{p}_{i_2}^k \\
\ket{01} & \vec{p}_{i_2}^k & 2\eta & \vec{p}_{i_1}^k + \vec{p}_{i_2}^k & \vec{p}_{i_1}^k + 2\vec{p}_{i_2}^k \\
\ket{10} & \vec{p}_{i_1}^k & \vec{p}_{i_1}^k + \vec{p}_{i_2}^k & 2\eta & 2\vec{p}_{i_1}^k + \vec{p}_{i_2}^k \\
\ket{11} & \vec{p}_{i_1}^k + \vec{p}_{i_2}^k & \vec{p}_{i_1}^k + 2\vec{p}_{i_2}^k & 2\vec{p}_{i_1}^k + \vec{p}_{i_2}^k & 2\eta
\end{array}
\end{equation}
Its graphical representation is shown in Figure \ref{fig:graph_op}.
To avoid clutter, we have not shown the overall multiplicative factor of $\frac{1}{n}$. The edge weights correspond to the off-diagonal elements of the perception operator, and node weights contribute to the uncertainty parameter in perception. For three neighbors, the computational basis for the density matrix is: \{$\ket{000}, \ket{001}, \ket{010}, \ket{011},\ket{100}, \ket{101}, \ket{110}, \ket{111}$\}. If $\alpha$ stands for $\ket{100}$  and $\beta$ for $\ket{101}$, $L_\alpha = (1,0,0)$, $L_\beta = (1,0,1)$ and $\vec{\Gamma}_i^k = (\vec{p}_{i_1}^k, \vec{p}_{i_2}^k, \vec{p}_{i_3}^k)$. Thus, the off-diagonal matrix element $u_{\alpha \beta}= 2\vec{p}_{i_1}^k+ \vec{p}_{i_3}^k$.

\subsection{Evolution of Perception Operator}
The expectation value for $O_i^k$ represents an internal perceptual average, making it a natural bridge to link perception with external motion. It is usually denoted as $\langle O_i^k \rangle$. Mathematically, it is expressed as 
\begin{equation}
\langle O_i^k \rangle = \mathrm{Tr}(\rho_i\, O_i^k),
\label{eq:o_avg}
\end{equation}
where the density matrix $\rho_i$ encodes the quantum nature of a perceptual decision state and governs the average value for the perception operator. We now introduce the most crucial step of connecting the perceptual Hilbert space to physical space. We hypothesize  $\langle O_i^k \rangle$ for the $k$-th axis to be proportional to the corresponding time derivative of the momentum $\vec{p}_i^k$ of the agent along the $k$-th axis. This means the average perception of an agent is the primary driving force behind the motion of the agents, like birds. In other words, we have
\begin{align}
    \frac{d \vec{p}_i^k}{d t}  &= \kappa \langle O_i^k \rangle
    \label{eq:oi}
\end{align}
where $\kappa$ is the proportionality constant that will be fixed by normalizing momentum directions. In Appendix section \ref{sec:trace}, we explicitly calculate that $\langle O_i^k \rangle$ for various entangled states and show that it contributes average momentum components of neighbors, analogous to the alignment rule in the Vicsek model \cite{vicsek1995novel}. By using the chain rule, we can express the time evolution of the perception operator $O_i^k$ as follows.
 
\begin{align}
\frac{d O_i^k}{dt} &= \frac{\partial O_i^k}{\partial t}  + \sum_{\gamma \in {\cal{N}}_i} \frac{\partial O_i^k}{\partial \vec{p}_{\gamma}^k}  \frac{\partial \vec{p}_{\gamma}^k}{\partial t}  \\
&=\kappa \sum_{\gamma \in {\cal{N}}_i} \langle O_\gamma^{k} \rangle  \frac{\partial O_i^k}{\partial \vec{p}_{\gamma}^k},
\label{eq:do_dt_full}
\end{align}
where ${\cal{N}}_i$ is the set of neighbors within the vision cone of angle $\alpha$ for agent $i$. Since the perception operator does not have explicit time dependence, we have dropped $\frac{\partial O_i^{k}}{\partial t}$ in the above equation. The matrix elements $O_i^k \big\lvert_{\alpha \beta}$ are as follows:
\begin{align}
    O_i^k \big\lvert_{\alpha \beta} &= \frac{1}{n} (\vec{L}_\alpha + \vec{L}_\beta)\cdot \vec{{\Gamma}}_i^k + \eta_i^k {\delta}_{\alpha \beta}, 
\end{align}
where ${\delta}_{\alpha \beta}$ is the Kronecker delta function. 
We find
\begin{align}
\frac{\partial O_i^k}{\partial \vec{p}_{\gamma}^k} \Big \lvert_{\alpha \beta} &= \frac{1}{n} \sum_{\xi \in {\cal{N}}_i}(\vec{L}_\alpha + \vec{L}_\beta)_\gamma \delta_{\xi \gamma}.
\label{eq:do_dpk}
\end{align}
Using equation \ref{eq:do_dpk} in equation \ref{eq:do_dt_full}, we obtain 
\begin{align}
    \frac{d O_i^{k}}{dt} \Big\lvert_{\alpha \beta} &= \frac{\kappa}{n} \sum_{\gamma \in {\cal{N}}_i} (\vec{L}_\alpha + \vec{L}_\beta)_\gamma \langle O_\gamma^k \rangle\\
    &=\frac{\kappa}{n} (\vec{L}_\alpha + \vec{L}_\beta)\cdot \vec{{\Theta}}_i^k,
    \label{eq:op_evolution1}
\end{align}
where $\vec{\Theta}_i^k$ is the $n$ dimensional vector formed by $\langle O_\gamma^k \rangle$'s of $n$ neighbors of $i$th agent. Thus, the evolution of the perception operator explicitly depends on the expectation value of the perception operator for neighbors. This essentially leads to a feedback effect and the source of non-linearity in the system.

Now, we need to discuss the perceptual Hamiltonian $X_i^k$ of the agent $\text{A}_i$. As argued before, the cognitive structure and subjective evaluation of each agent are considered independent. This leads to independent decision-making for every agent. Thus, the perceptual Hilbert space of each agent is disjoint, and the system is non-dissipative in perceptual dynamics. Thus, it follows the unitary evolution of the perception operator. Consequently, the evolution of the perception operator in the Heisenberg picture can be expressed as follows:
\begin{align}
    \frac{d O_i^{k}}{dt}  &= \frac{i}{\hbar} [X_i^k,O_i^k]. 
    \label{eq:do_dt_2}
\end{align}
It is worth mentioning that the physical space dynamics of self-propelled active agents is a non-equilibrium and dissipative one. Thus, one cannot define a Hamiltonian in physical space. However, our framework provides a complementary interpretation by describing its perceptual evolution as a conservative system.
\subsection{Connection with physical space momenta}

Equation \ref{eq:op_evolution1} gives us the matrix elements of the time derivative of the perception operator, and equation \ref{eq:do_dt_2} provides the quantum-like dynamical evolution of the operator $O_i^k$. It is worth reminding that the off-diagonal elements of the perception matrix $O_i^k$ (see equation \ref{eq:o_matrix}) are defined through the physical space momenta of neighbors in the vision cone. From equation \ref{eq:op_evolution1}, we learn that the evolution of the perception operator for agent $i$ depends on the expectation value of the perception operator for all neighbors $\gamma \in \mathcal{N}_i$. 

One interesting case arises here. As per the previous discussion, if two agents share a common neighbor, how does the dynamics of the perception operator evolve?
From equation \ref{eq:op_evolution1}, we infer that the expectation value of this common neighbor's perception operator contributes to both agents who perceive it as a neighbor. On the other hand, this common neighbor's own neighborhood structure decides its state, and momenta dictate the form of the perception operator. Thus, we can observe a feedback of physical space momenta through the inherent structure of the perception operator.

At every time step, we know the physical space momenta and relative position of agents. It defines the vision cone for each agent, thereby determining the corresponding perception operator and the density matrix that characterizes the agent’s perceptual decision state. Thus, the perceptual Hamiltonian can be obtained by solving for $X_i^k$ in equation \ref{eq:do_dt_2}. To solve it, we need to use the vectorization (or vec) operator method \cite{magnus2007matrix} (see Appendix section \ref{sec:vec}). As mentioned earlier, we cannot define a physical space Hamiltonian for such a system. The agents have self-propulsion, and they constantly take in energy from the environment or internal system (e.g., metabolism, battery, etc.) and convert it into motion. So far as kinetic energy is concerned, the total kinetic energy $E=\sum_{i=1}^N \sum_{k=1}^{d} \frac{\left ({p_i^k}\right )^2}{2}$ is constant since we are considering constant speed for the agents. 

Thus, we understand that although physical space dynamics is non-conservative (non-unitary), the perceptual Hamiltonian evolves with time unitarily and contributes to flocking behavior in physical space. The order parameter $\langle \phi_v \rangle$ quantifies the average alignment of agent momentum $\vec{p}_i $, indicating the level of collective order in the system. This is the most important quantity so far as collective motion is concerned. It is defined as:
\begin{equation} 
\langle \phi_v \rangle = \frac{1}{Nv_0} \left | \sum_{i=1}^{N} \vec{p}_i \right |,
\end{equation} 
where $v_0$ is the constant speed of the agents.
Next, we introduce two important measures from perceptual space dynamics that closely mimic the order parameter of the system in physical space. 

\subsection{Perceptual measures of flocking}
\label{sec:perceptual_measures}
For a set of eigenvalues $\{\omega_i^k\}$ of perceptual Hamiltonian $X_i^k$, we define perceptual energy $\cal{E}$ as follows.
\begin{align}
    \cal{E} &= \sum_{i=1}^{N}  \sum_{k=1}^{d}  \text{max}[\{\omega_i^k\}],
    \label{eq:energy}
\end{align}
where perceptual energy \( \mathcal{E} \) can be intuitively understood as a measure of the overall cognitive readiness of a multi-agent system. It aggregates the peak perceptual evaluations across all agents and directions, reflecting how easily the system, as a whole, can perceive and react to environmental conditions. A higher value of \( \mathcal{E} \) probably suggests that agents require stronger cognitive functioning to work towards order and cohesion in the flock. In other words, the lower value of \( \mathcal{E} \) makes it easier to respond to their perceptual inputs. Therefore, as the system becomes more ordered, $\mathcal{E}$ reduces since the perceptual effort needed to sustain flocking behavior diminishes, as cohesive motion is more easily achieved. This effect will be demonstrated in our subsequent numerical simulations.

For a set of eigenvalues of the operator $O_i^k$ given by a set $\{\epsilon_i^k\}$, we define $\text{max}[\{\epsilon_i^k\}]$ as the $k$-th component of the perception vector and its mean Euclidean norm as the perception strength $\cal{P}$ at a particular time step. In other words,  
\begin{align}
    \cal{P} &= \left \| \sum_{i}^{N}  \sum_{k=1}^{d}  \text{max}[\{\epsilon_i^k\}] \hat{e}_k\right \|,
    \label{eq:perception}
\end{align}
$\hat{e}_k$ is the unit vector along $k$th direction and $d$ is the dimension of the periodic box. Perception strength \( \mathcal{P} \) quantifies the degree of directional coherence in the perceptual field of a multi-agent system. It reflects how strongly agents align with the flock based on their perceptual evaluations across spatial dimensions. A high value of \( \mathcal{P} \) corresponds to coherent, flock-like behavior, while a low \( \mathcal{P} \) indicates disordered phases. Our subsequent numerical simulations reveal that this perceptual measure exactly parallels the order parameter defined earlier.

The parameter \( \eta \), which enters the diagonal elements of the perception operator, acts as a tuning variable that modulates the agents’ perceptual uncertainty. As \( \eta \) increases, perceptual decisions become more randomized, reducing coherence and driving the system toward a disordered phase. Conversely, low \( \eta \) enhances directional alignment, promoting the emergence of ordered, collective dynamics. This behavior is closely linked to noise-induced phase transitions in the system, as often discussed in connection with the Vicsek model \cite{vicsek1995novel} for flocking.

\section{Perceptual basis of Vicsek model}
\label{sec:vicsek}

As stated earlier, one of the focuses of this work is to derive the classical Vicsek model of flocking from the perceptual dynamics. For completeness, we now mention the Vicsek model \cite{vicsek1995novel}. This classical nonlinear model captures the emergence of collective motion in a system of self-propelled agents. In this model, the momentum of the $i$th agent $\vec{p}_i$ at time $(t+1)$ is expressed as follows.
\begin{equation}
\vec{p}_i(t+1) = \vec{p}_i (t)+ \frac{1}{n} \sum_{j \in \mathcal{N}_i} \vec{p}_j(t) + \eta v_0\hat{e}_i(t),
\label{eq:vicsek}
\end{equation}
where $\mathcal{N}_i$ is the set of $n$ neighboring agents, $v_0$ is the constant speed of an agent, $\eta$ is the noise strength and $\hat{e}_i(t)$ is a random unit vector. The primary component is the average momentum of neighbors. Different variants of the Vicsek model essentially differ in terms of their neighborhood selection algorithms.

At each time step, an agent updates its momentum direction by aligning with the average momentum of its neighbors, reflecting a local tendency toward consensus. This alignment mechanism models the fundamental principle of imitation or local influence observed in natural swarms. The second term introduces stochasticity through a random vector scaled by the noise strength \( \eta \), which accounts for uncertainty in perception or fluctuations in decision-making. As a result, the model exhibits a noise-induced phase transition: low \( \eta \) leads to global order with agents moving coherently, while high \( \eta \) results in disordered, random motion. The interplay between local alignment and noise encapsulates the key ingredients for the self-organization of collective behavior.

Now, we wish to demonstrate that our proposed framework essentially boils down to a Vicsek-like dynamics in physical space for certain entangled states.
For example, let's take the case of the $\ket{\Phi^+}=\frac{1}{2}(\ket{00}+\ket{11})$ state. The equation \ref{eq:oi} becomes as follows:
\begin{align}
\frac{d \vec{p}_i^k}{d t} &=  \kappa \langle O_i^{k} \rangle ,\\
&=  \frac{\kappa}{2}(\vec{p}_{i_1}^k+\vec{p}_{i_2}^k) + \kappa \, \eta.
\end{align}
Thus, by collecting spatial dimension components, we obtain
\begin{align}
\frac{d \vec{p}_i}{d t} &= \kappa \left[\frac{1}{2}(\vec{p}_{i_1}+\vec{p}_{i_2}) + \sum_{k=1}^{d} \eta \hat{e}_i^k \right].
\end{align}
With discrete time steps, we can write
\begin{align}
\vec{p}_i (t+1)& = \vec{p}_i (t)+ \kappa \left[\frac{1}{2}(\vec{p}_{i_1}+\vec{p}_{i_2}) + \sum_{k=1}^{d} \eta \hat{e}_i^k \right].
\end{align}
Comparing this equation with equation \ref{eq:vicsek}, we can observe a striking parallel with the Vicsek model and the $\eta$ term mimicking the noise parameter.

For a uniform superposed state $\ket{\psi}=\frac{1}{2}(\ket{00}+\ket{01}+\ket{10}+\ket{11})$, physical momentum at $(t+1)$ can be expressed as follows:
\begin{align}
\vec{p}_i (t+1)& = \vec{p}_i (t)+ \kappa \left[\frac{3}{2}(\vec{p}_{i_1}+\vec{p}_{i_2}) + \sum_{k=1}^{d} \eta \hat{e}_i^k \right].
\end{align}
The evolutions described above closely resemble the Vicsek model, where dynamics are governed by the average momentum of neighbors combined with a noise term. Although we have considered specific entangled or superposed states, our numerical results demonstrate that this behavior generalizes to any generic state with positive coefficients. Negative coefficients, such as those in the $\Phi^-$ state, signify perceptual conflict between elementary choices. Such a phase difference acts destructively. Our numerical simulation does not yield order or cohesion in these cases. Thus, we have not discussed these states in the subsequent section on numerical results. 

\begin{figure}[!t]
    \centering
    \subfloat[]{\includegraphics[width=0.45\linewidth]{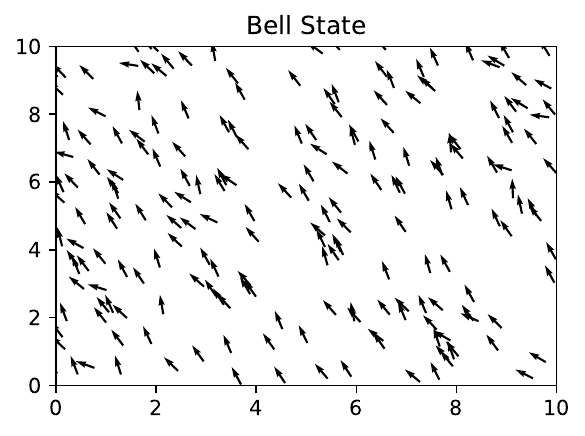}}
    \subfloat[]{\includegraphics[width=0.45\linewidth]{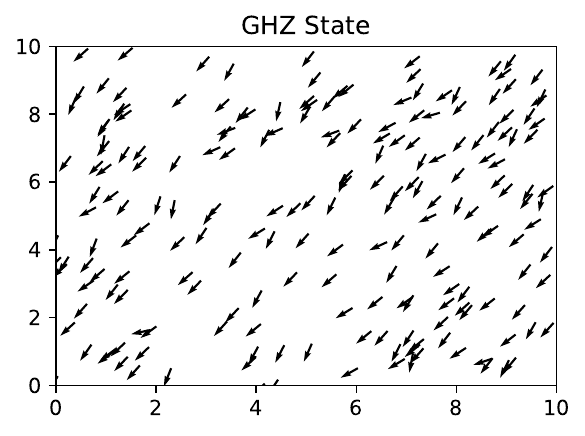}}
    \caption{Flocking patterns for (a) $\Phi^+$and (b) $\text{GHZ}_3$ states at noise strength $\eta=0.2$ and $t=100$. }
    \label{fig:pattern}
\end{figure}

The destructive effect of the $\Phi^-$ state can be shown mathematically as follows. Applying the same procedure as above, we obtain the following dynamics for the $\Phi^-$  state.
\begin{align}
\vec{p}_i (t+1)& = \vec{p}_i (t)+ \kappa \left[-\frac{1}{2}(\vec{p}_{i_1}+\vec{p}_{i_2}) + \sum_{k=1}^{d} \eta \hat{e}_i^k \right].
\end{align}
Clearly, the $\Phi^-$ state yields a dynamics wherein average neighbor momenta contribute destructively. The situation remains the same for the $\Psi^-$ or a generic superposed state with a random negative coefficient. This is a characteristic departure from classical models for collective motion, such as the Vicsek model.

\section{Numerical simulations}
\label{sec:numerical}

We simulate \( N = 200 \) agents of unit mass in a 2D periodic box of $L=10$ for \( 10^3 \) time steps. The initial position and the momentum directions are chosen randomly. The distance between agents is measured using the standard Euclidean distance in the periodic box, and the agents' speed, \( v_0 \), is fixed at 0.5. The evolution of momenta is considered with a time step of $\Delta t=0.1$. The noise strength $\eta$ is fixed at 0.2. This moderate $\eta$ is chosen to demonstrate that the system attains significant order even in the presence of noise \cite{beuriaricci2024}. However, we also choose $0 \leq \eta \leq 2.0$ in steps of 0.1 while studying the noise-induced phase transition. For $\eta\geq1.0$, the perceptual uncertainty overpowers the alignment, and the system becomes extremely disordered. 

We consider a vision cone with $\alpha=\frac{\pi}{2}$, $r_{min} = 0.1$, and  $r_{max} = 5$. Although we report these representative values for the radial region of perception, the results do not change qualitatively for other choices. $r_{min}>0$ represents the so-called non-local model \cite{beuriaricci2024}, as nearby agents do not 
 explicitly influence. For numerical calculations, we have discretized $\frac{d \vec{p}_i^k}{d t}$ in equation \ref{eq:oi} as the difference between $\vec{p}_i^k$'s of two successive time steps divided by step size $\Delta t$. Thus, the values of $\vec{p}_i^k$ at time $t+1$ are obtained in terms of the expectation values of perception operators and momentum components of agents at $t$. 

In addition to the entangled states for \(n = 2\) and \(n = 3\) discussed earlier, we also consider their corresponding uniform and random superposition states. For the random superposition case, the state evolves at each time step by randomly selecting the coefficients in equation \ref{eq:random_sup}, with an analogous procedure applied for the \(n = 3\) state. In Figure \ref{fig:pattern}, we present a typical flocking pattern for two entangled states, namely, $\Phi^+$ and $\text{GHZ}_3$ states at noise strength $\eta=0.2$ and $t=100$.

\begin{figure}[!ht]
    \centering
    \subfloat[\label{fig:op_90a}]{\includegraphics[width=0.45\linewidth]{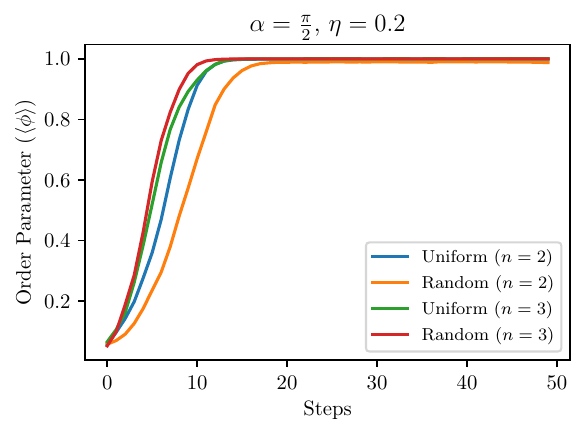}}
    \subfloat[\label{fig:op_90b}]{\includegraphics[width=0.45\linewidth]{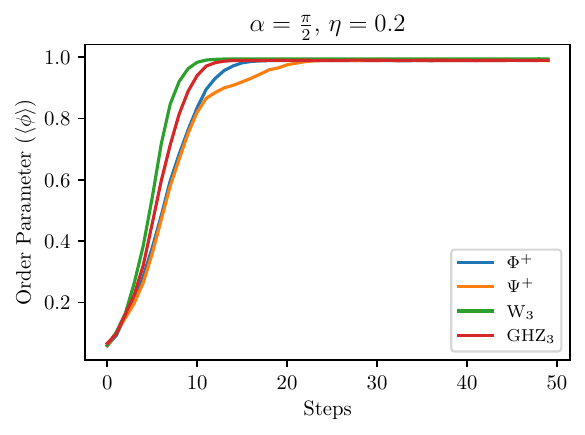}}
    \caption{Order parameter vs. time steps at $\eta = 0.2$: (a) Uniform and random superposition states; (b) Maximally entangled states for $n=2$ and $n=3$.}
    \label{fig:op_90}
\end{figure}

\begin{figure}[!ht]
    \centering    
    \subfloat[\label{fig:op_etaa}]{\includegraphics[width=0.45\linewidth]{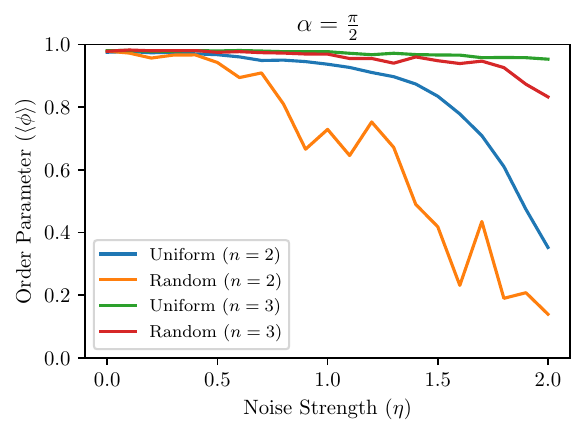}}
    \subfloat[\label{fig:op_etab}]{\includegraphics[width=0.45\linewidth]{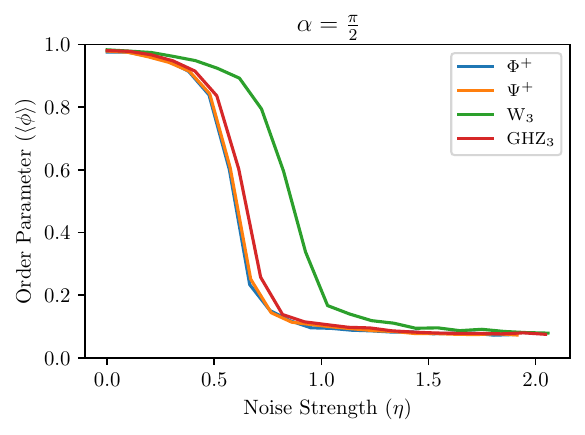}}
    \caption{Order parameter vs. noise strength $\eta$: (a) Uniform and random superposition states; (b) Maximally entangled states for $n=2$ and $n=3$.}.
    \label{fig:op_eta}
\end{figure}

\begin{figure}[!ht]
    \centering
    \subfloat[\label{fig:perceptiona}]{\includegraphics[width=0.45\linewidth]{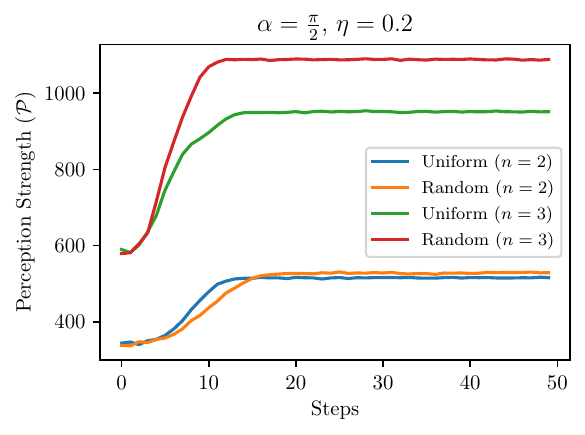}}
    \subfloat[\label{fig:perceptionb}]{\includegraphics[width=0.45\linewidth]{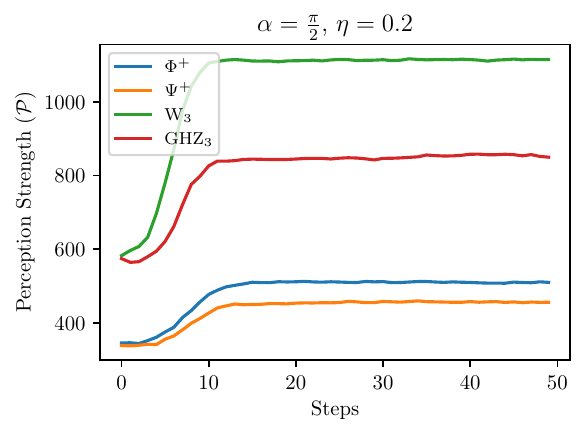}}
    \\
    \subfloat[\label{fig:perceptionc}]{\includegraphics[width=0.45\linewidth]{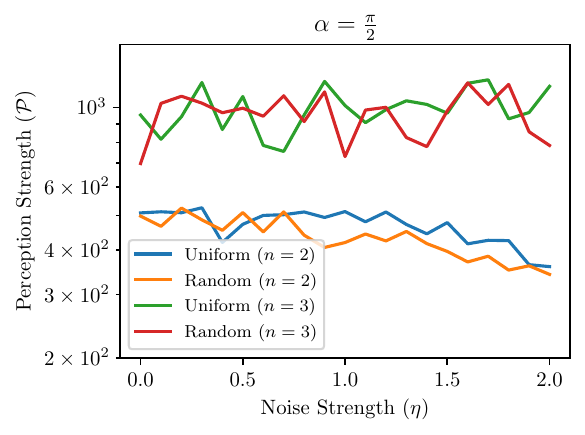}}
    \subfloat[\label{fig:perceptiond}]{\includegraphics[width=0.45\linewidth]{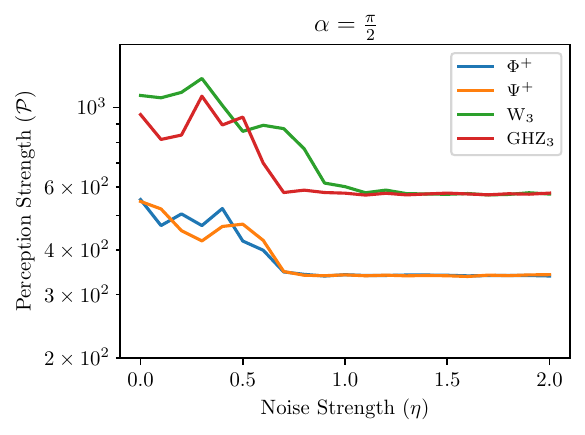}}

\caption{Variation of perception strength: (a)–(b) Over time for uniform, random superposition, and maximally entangled states at $\eta = 0.2$; (c)–(d) Time-averaged values versus noise strength $\eta$ for the same state types.}
    \label{fig:perception}
\end{figure}

\begin{figure}[!ht]
    \centering
    \subfloat[\label{fig:energya}]{\includegraphics[width=0.45\linewidth]{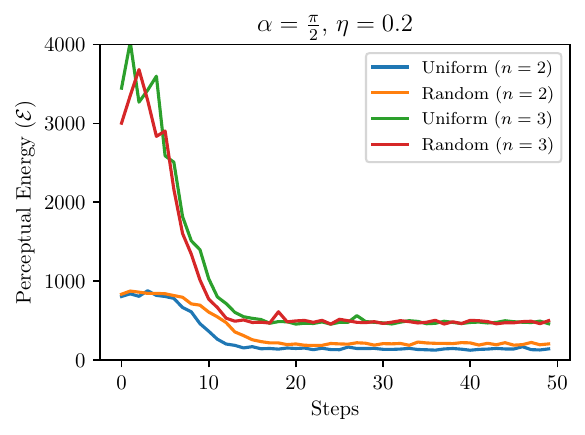}}
    \subfloat[\label{fig:energyb}]{\includegraphics[width=0.45\linewidth]{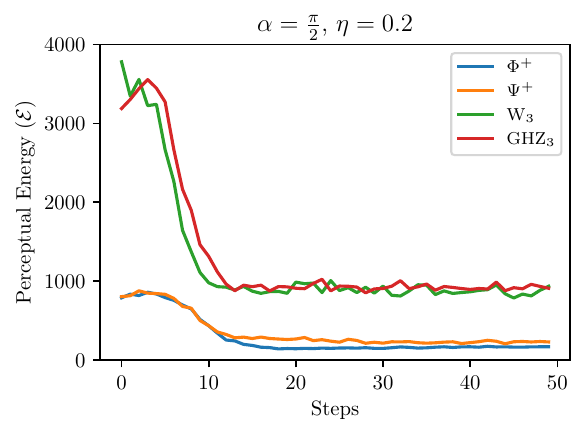}}
    \\
    \subfloat[\label{fig:energyc}]{\includegraphics[width=0.45\linewidth]{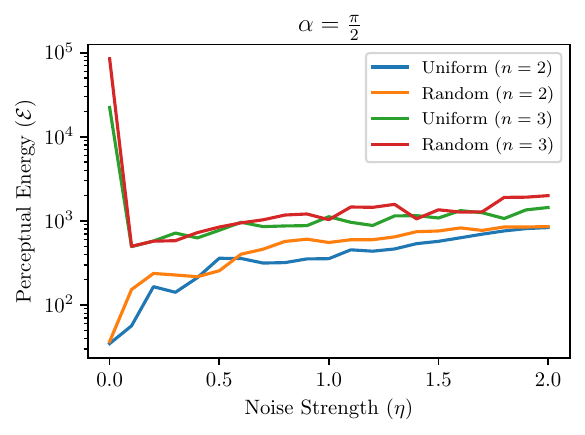}}
    \subfloat[\label{fig:energyd}]{\includegraphics[width=0.45\linewidth]{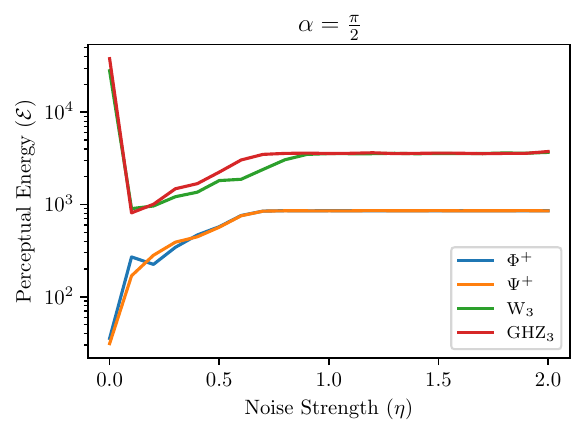}}
    \caption{Variation of perceptual energy: (a)–(b) Over time for uniform, random superposition, and maximally entangled states at $\eta = 0.2$; (c)–(d) Time-averaged values versus noise strength $\eta$ for the same state types.}
    \label{fig:energy}
\end{figure}

In Figure~\ref{fig:op_90}, we present the variation of order parameter $\langle \phi_v \rangle$ with time at a fixed noise strength $\eta=0.2$ and $\alpha=\frac{\pi}{2}$. Figure~\ref{fig:op_90a} presents the uniform and random superposition states for $n=2$ and $n=3$. Figure~\ref{fig:op_90b} corresponds to the maximally entangled Bell states $\Psi^+$ and $\Phi^+$ as well as the $\text{GHZ}_3$ and $\text{W}_3$ states. We observe in both plots that the $n=3$ state attains order a bit quickly compared to the $n=2$ state. This can be attributed to enhanced perception in the $n=3$ case.

In Figure \ref{fig:op_eta}, we present the typical phase transition induced by the noise strength $\eta$. In other words, an ordered flock breaks away as the noise increases. The situation is quite similar to the Vicsek-like models \cite{beuriaricci2024}. Comparing Figure \ref{fig:op_etaa} and \ref{fig:op_etab}, we understand that the entangled states become ineffective at moderate to large noise strength. On the contrary, the superposed states, particularly the three-neighbor scenarios, maintain flock even in the presence of very large noise. This can be appreciated as follows. In the low-noise case, the interdependent decision-making is much easier compared to the high-noise case. High noise supports only the superposed states, implying that independent individual decision-making for each perceived neighbor still remains a viable mechanism. Also, in a high noise scenario, the uniform superposition yields a larger order than its random superposition states (see Figure \ref{fig:op_etaa}).

So far, we have discussed the evolution of the order parameter with time steps and noise strengths. Now we focus on two perceptual measures, namely, perception strength and perceptual energy (see section \ref{sec:perceptual_measures}). In Figure~\ref{fig:perception}, we present the overall perception strength $\cal{P}$ of a flock as defined in equation \ref{eq:perception}. The first (second) row corresponds to the variation of perception strength $\cal{P}$ with time (noise strength). In all plots, we observe that the perception strength for $n=2$ states is lower than for $n=3$ states. This is intuitive because more neighbors demand more perceptual evaluation. 

In Figure \ref{fig:perceptiona}, although $n=2$ uniform and random superposition states saturate to similar values, the gap between these states for the $n=3$ random superposition case is significantly larger. The situation is similar in the case of maximally entangled states presented in Figure \ref{fig:perceptionb}. Another important point to note is that the variation of the perception strength $\mathcal{P}$ closely mirrors the behavior of the order parameter $\langle \phi \rangle$ shown in earlier plots. This highlights $\mathcal{P}$ as a meaningful perceptual analogue to the traditional order parameter. The saturation of $\mathcal{P}$ after the flock attains order suggests that perception strength stabilizes once collective alignment is achieved.

In the second row, we present the variation of the time-averaged perception strength as a function of the noise parameter $\eta$. In addition to the hierarchy observed for $n=2$ and $n=3$ states in the first row, we find that for maximally entangled states (in Figure \ref{fig:perceptiond}), the perception strength $\mathcal{P}$ decreases with increasing $\eta$ and eventually saturates for sufficiently large $\eta$. This trend once again parallels the behavior of the order parameter $\langle \phi \rangle$. However, $\mathcal{P}$ exhibits minimal change across the chosen $\eta$ range in Figure \ref{fig:perceptionc}. This indicates that maximally entangled states are more fragile under high noise conditions, reflecting their reduced robustness to contribute to the flocking.

In Figure~\ref{fig:energy}, we present the variation of perceptual energy $\cal{E}$, another important measure of cohesion in the flock. The first row illustrates the evolution of $\mathcal{E}$ over time at a fixed noise level $\eta = 0.2$, while the second row displays the time-averaged values of $\mathcal{E}$ as a function of the noise strength $\eta$. The time evolution of superposed states in Figure \ref{fig:energya} behaves pretty similarly to the maximally entangled states in Figure \ref{fig:energyb}. After certain steps, the perceptual energy saturates to a low value, marking the collective perceptual ease in maintaining the flock once cohesion has been achieved. We also observe that $n=3$ has a larger $\cal{E}$ compared to $n=2$. This is again intuitive because a larger number of neighbors demands a stronger effort in perceptual decision-making. 

In Figure \ref{fig:energyc} and \ref{fig:energyd}, we present the corresponding variation of time-averaged perceptual energy with noise strength $\eta$. We observe that the perceptual energy $\mathcal{E}$ tends to saturate at a constant value under high noise conditions, as seen in Figure~\ref{fig:energyd}. Large noise strength corresponds to a state of complete disorder in the flock, making it a perceptually demanding scenario for the agents. However, the behavior differs slightly for the superposed states shown in Figure~\ref{fig:energyc}, where the average $\mathcal{E}$ continues to increase even at high $\eta$. This aligns with our earlier observations on the order parameter, suggesting that these superposed states can maintain a significant degree of cohesion within the flock, even in the presence of strong noise. Thus, $\mathcal{E}$ does not drop drastically in these cases.

\section{Conclusion}
\label{sec:conclusion}
In this work, we have studied the perceptual decision-making of agents leading to cohesion in the flock by leveraging the framework of quantum dynamics. Although we have considered visual perception, considering the flocking of birds, the developed framework applies to any multi-agent system with a generic notion of perception of neighbors. Unlike the traditional approach of studying the flocking through classical nonlinear models such as the Vicsek model, we suggest that superposition and entanglement in individual perceptual decision states drive the dynamics in the physical space. In particular, we have explored the uniform and random superposition states as well as maximally entangled states such as Bell, GHZ, and W states for two and three neighbors for each agent.

We have also introduced a perception operator encoding the perceptual evaluations associated with the decision-making process. The perception operator can be described as a fully connected graph where nodes represent the decision basis states. We have proposed that the force on an agent in physical space is governed by the quantum average of its perception operator. This hypothesis gives rise to the Vicsek-like model in physical space. 

Our work also reveals the rich nonlinear dynamics that emerge from considering different perceptual decision states. In low-noise regimes, both superposition and entangled states lead to substantial cohesion in the flock, as reflected in the high value of the order parameter. However, under high noise conditions, maximally entangled states fail to sustain significant order in the flock, in contrast to the more robust behavior of uniform and random superposition states. This distinction is difficult to capture within classical nonlinear models, highlighting the potential advantages of adopting a quantum framework for modeling collective decision-making.

We also suggest two important quantum measures of collective behavior: perception strength and perceptual energy as the perceptual analogue of the order parameter. Although we have chosen a very specific kind of perception operator motivated by the popular Vicsek model, the framework can be extended to other novel scenarios of collective motion. This exploratory study opens up new possibilities in studying perceptual dynamics of collective behavior, which otherwise may not be well appreciated in classical systems. Our future works will further explore the dynamics of perceptual decision states contributing to rich collective motion.

\enlargethispage{20pt}



\vskip2pc

\bibliographystyle{elsarticle-num}

\bibliography{ref}

\appendix
\renewcommand{\thesection}{\Alph{section}} 
\numberwithin{equation}{section}           
\numberwithin{figure}{section}             
\numberwithin{table}{section}              

\section{Different entangled states}
\label{sec:trace}

For two qubits, we consider the maximally entangled Bell states as follows:
\begin{align}
    |\Phi^+\rangle &= \frac{1}{\sqrt{2}} (|00\rangle + |11\rangle),
    |\Phi^-\rangle = \frac{1}{\sqrt{2}} (|00\rangle - |11\rangle),\nonumber \\
    |\Psi^+\rangle &= \frac{1}{\sqrt{2}} (|01\rangle + |10\rangle),
    |\Psi^-\rangle = \frac{1}{\sqrt{2}} (|01\rangle - |10\rangle)
    \label{eq:trace_bell}
\end{align}

For three qubits, some well-known entangled states are the Greenberger-Horne-Zeilinger (GHZ) state and W states given as follows:
\begin{align}
    \ket{\text{GHZ}_3} &= \frac{1}{\sqrt{2}} \left(\ket{000} + \ket{111}\right), \nonumber \\
    \ket{\text{W}_3} &= \frac{1}{\sqrt{3}} \left( \ket{001} + \ket{010} + \ket{100} \right).
    \label{eq:trace_wghz}
\end{align}

\begin{figure}[!t]
    \centering
    \includegraphics[width=0.8\linewidth]{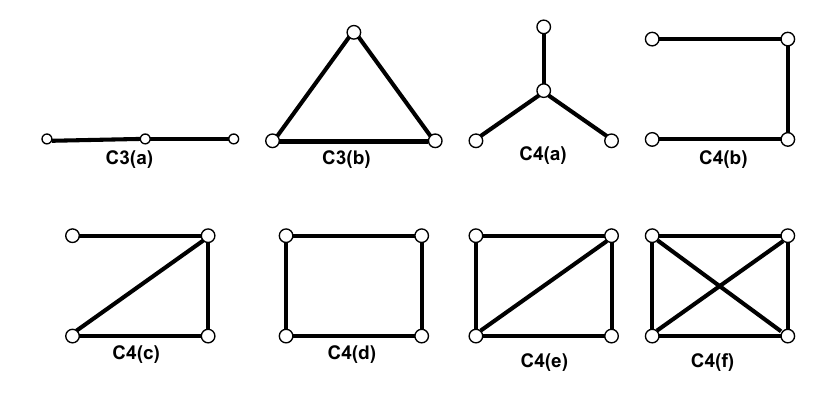}
    \caption{Cluster states from connected graphs with $n$=3 and $n=4$ nodes.}
    \label{fig:cluster-states}
\end{figure}

Another class of highly entangled states is called cluster states, which are defined on graphs. 
Let \( G = (V, E) \) be a graph on \( n \) vertices and E edges. To define an \( n \)-qubit cluster state on graph \( G \), every node is represented by $\ket{+}=\frac{\ket{0}+\ket{1}}{\sqrt{2}}$ state. Thus, the cluster state $\ket{\psi_G} \in (\mathbb{C}^2)^{\otimes n}$ is given by
\begin{equation}
|\psi_G\rangle = \prod_{e_{ij} \in E} (CZ)_{ij} |+\rangle^{\otimes n},
\label{eq:cluster}
\end{equation}
where $e_{ij}$ is an edge and  \( (CZ)_{ij} \) denotes the C-phase gate applied on the edge connecting qubits \( i \) and \( j \). We also have 
\begin{equation}
    CZ|ab\rangle = (-1)^{ab} |ab\rangle,
\end{equation}
where \( a, b \in \{0,1\} \). It is also to be noted that various \( (CZ)_{ij} \) commute with each other, so the product in equation \ref{eq:cluster} is well-defined. In Figure \ref{fig:cluster-states}, we present connected graphs for $n=3$ and $n=4$ nodes for constructing cluster states. 

In order to find the updated momentum orientation of an agent, we need to know the expectation values of operators in equation \ref{eq:oi}. Thus, we calculate the $\mathrm{Tr}(\rho_i O_i^{A_k})$ values for various entangled states in terms of momentum components of neighbors for agent $i$. For a Bell state, we have two neighbors represented by $i_1$ and $i_2$. The trace operation gives
\begin{align}
    \mathrm{Tr}(\rho_i O_i^{A_k})=
\begin{cases} 
\frac{1}{2}(\vec{p}_{i_1}^k+\vec{p}_{i_2}^k)+\eta  & \text{if } \rho_i =\ket{\Phi^+}\bra{\Phi^+} \\
-\frac{1}{2}(\vec{p}_{i_1}^k+\vec{p}_{i_2}^k)+\eta  & \text{if } \rho_i =\ket{\Phi^-}\bra{\Phi^-} \\
\frac{1}{2}(\vec{p}_{i_1}^k+\vec{p}_{i_2}^k)+\eta  & \text{if } \rho_i =\ket{\Psi^+}\bra{\Psi^+} \\
-\frac{1}{2}(\vec{p}_{i_1}^k+\vec{p}_{i_2}^k)+\eta  & \text{if } \rho_i =\ket{\Psi^-}\bra{\Psi^-}
\end{cases}
\label{eq:tr_bell}
\end{align}
For $\text{GHZ}_3$ and $\text{W}_3$ states, we obtain as follows.
\begin{align}
    \mathrm{Tr}(\rho_i O_i^{A_k})=
\begin{cases} 
\frac{1}{3}(\vec{p}_{i_1}^k+\vec{p}_{i_2}^k + \vec{p}_{i_3}^k) +\eta & \text{if } \rho_i =\ket{\text{GHZ}_3}\bra{\text{GHZ}_3} \\
\frac{9}{4}(\vec{p}_{i_1}^k+\vec{p}_{i_2}^k + \vec{p}_{i_3}^k)+\eta  & \text{if } \rho_i =\ket{\text{W}_3}\bra{\text{W}_3}
\end{cases}
\label{eq:tr_wghz}
\end{align}
The trace operation for 3-qubit cluster states yields as follows.
\begin{align}
    \mathrm{Tr}(\rho_i O_i^{A_k})=
\begin{cases} 
\frac{1}{3}(-\vec{p}_{i_1}^k+\vec{p}_{i_2}^k + \vec{p}_{i_3}^k) +\eta & \text{if } \rho_i =\ket{\psi_{_{\text{C3(a)}}}}\bra{\psi_{_{\text{C3(a)}}}} \\
-\frac{1}{3}(\vec{p}_{i_1}^k+\vec{p}_{i_2}^k + \vec{p}_{i_3}^k) +\eta & \text{if } \rho_i =\ket{\psi_{_{\text{C3(b)}}}}\bra{\psi_{_{\text{C3(b)}}}}
\end{cases}
\label{eq:tr_c3}
\end{align}
Similarly, the 4-qubit cluster states give  $\mathrm{Tr}(\rho_i O_i^{A_k})$
\begin{align}
=
\begin{cases} 
\frac{1}{4}(-\vec{p}_{i_1}^k+3\vec{p}_{i_2}^k + 3\vec{p}_{i_3}^k+ 3\vec{p}_{i_4}^k) +\eta & \text{if } \rho_i =\ket{\psi_{_{\text{C4(a)}}}}\bra{\psi_{_{\text{C4(a)}}}} \\
-\frac{1}{4}(\vec{p}_{i_1}^k+\vec{p}_{i_2}^k + \vec{p}_{i_3}^k + \vec{p}_{i_4}^k)+\eta  & \text{if } \rho_i =\ket{\psi_{_{\text{C4(b)}}}}\bra{\psi_{_{\text{C4(b)}}}} \\
-\frac{1}{4}(\vec{p}_{i_1}^k+\vec{p}_{i_2}^k + \vec{p}_{i_3}^k - \vec{p}_{i_4}^k )+\eta  & \text{if } \rho_i =\ket{\psi_{_{\text{C4(c)}}}}\bra{\psi_{_{\text{C4(c)}}}} \\
\frac{3}{4}(\vec{p}_{i_1}^k+\vec{p}_{i_2}^k + \vec{p}_{i_3}^k + \vec{p}_{i_4}^k)+\eta  & \text{if } \rho_i =\ket{\psi_{_{\text{C4(d)}}}}\bra{\psi_{_{\text{C4(d)}}}} \\
-\frac{1}{4}(\vec{p}_{i_1}^k+\vec{p}_{i_2}^k + \vec{p}_{i_3}^k + \vec{p}_{i_4}^k) +\eta & \text{if } \rho_i =\ket{\psi_{_{\text{C4(e)}}}}\bra{\psi_{_{\text{C4(e)}}}} \\
\frac{1}{4}(\vec{p}_{i_1}^k+\vec{p}_{i_2}^k + \vec{p}_{i_3}^k+ \vec{p}_{i_4}^k)+\eta  & \text{if } \rho_i =\ket{\psi_{_{\text{C4(f)}}}}\bra{\psi_{_{\text{C4(f)}}}}
\end{cases}
\label{eq:tr_c4}
\end{align}

\section{Vectorization method}
\label{sec:vec}

Vectorization is a transformation that converts a matrix into a column vector by stacking its columns on top of one another. By using this vectorization method, commutator $[X, O]$ can be rewritten as a linear operation on $\text{vec}(X)$:

\begin{align}
    \text{vec}(\frac{d O_i^{k}}{dt})  &=\frac{i}{\hbar} \text{vec}([X_i^k, O_i^k]) \\
    &= \frac{i}{\hbar} ({O_i^k}^T \otimes I - I \otimes O_i^k) \, \text{vec}(X_i^k) \\
    \implies \text{vec}(X_i^k) &=-i\hbar ({O_i^k}^T \otimes I - I \otimes O_i^k)^{+} \text{vec}(\frac{d O_i^{k}}{dt}), 
    \label{eq:op_pseudo_inv}
\end{align}
where $({O_i^k}^T \otimes I - I \otimes O_i^k)^{+}$ corresponds to the Moore-Penrose pseudo inverse. We can reshape $\text{vec}(X_i^k)$ in equation \ref{eq:op_pseudo_inv} to obtain the Hamiltonian matrix of the system. Throughout this study, we prefer to work with the unit of $\hbar=1$. 

\section{Geometric measures of entanglement}

\begin{table*}[!ht]
\centering
\begin{tabular}{|c|c|c|c|c|c|c|c|c|c|c|c|}
\hline

& $\Psi^{\pm}$, $\Phi^{\pm}$ &  $\text{W}_3$ & $\text{GHZ}_3$ &  C3(a) & C3(b) & C4(a) & C4(b) & C4(c) & C4(d) & C4(e) & C4(f)  \\ \hline 
  
GME & 0.5 & 0.56 & 0.53  &  0.52 & 0.52 & 0.57 & 0.77 & 0.77 & 0.76 & 0.77 & 0.56\\ \hline
GMC & 0.5 & 0.67 & 0.5  & 0.87 & 0.87 & 0.94 & 0.94 & 0.94 & 0.94 & 0.94 & 0.94 \\ \hline
\end{tabular}
\caption{Geometric measure of entanglement (GME) and geometric measure of coherence (GMC) calculated numerically for different entangled states.}
\label{tab:gm}
\end{table*}

\label{sec:gme}
The geometric measure of entanglement (GME) \cite{streltsov2010linking, zhang2020numerical} is defined as:
\begin{equation}
\text{GME} = 1 - \max_{\sigma \in \text{SEP}} \, \text{F}(\rho, \sigma)^2,
\end{equation}
where \(\text{F}(\rho, \sigma)\) is the fidelity between the density matrix \(\rho\) and the separable state \(\sigma\). The fidelity is given by:
\begin{equation}
\text{F}(\rho, \sigma) = \text{Tr}  \sqrt{\sqrt{\sigma} \rho \sqrt{\sigma}}.
\end{equation}
Here, the maximization is performed over all separable states \(\sigma\).
The coherence of a quantum state \(\rho\) can be defined in terms of its fidelity with a maximally coherent state \(\rho_{\text{max}}\). The coherence measure \cite{streltsov2015measuring, zhang2020numerical}  is given by:
\begin{equation}
\text{GMC} = 1 - \max_{\sigma \in \mathcal{I}} \, \text{F}(\rho, \sigma)^2,
\end{equation}
where $\mathcal{I}$ is the set of all possible incoherent states. In Table \ref{tab:gm}, we have calculated  $\text{GME}$ and $\text{GMC}$ numerically for all the entangled states under consideration.

\end{document}